\input harvmac

%
\def\lb       {\left( }
\def\rb       {\right) }

\def\lbb     {\left[ }
\def\rbb      {\right] }
\def\comma      { \, , }
\def\period     { \, . }
\def\bra#1      { \langle \, #1 \, \vert \, }
\def\ket#1      { \, \vert \, #1 \, \rangle  }
\def\semiket#1  { \, #1 \, \rangle \, }
\def\del        {  \partial  }
\def\zbar       { \bar{z} }
\def\abs#1      {  \, \vert #1 \vert \,   }
\def\calF       { {\cal F} }
\def\calG       { {\cal G} }
\def\calN    { {\cal N} }
%
%

\def\xxx#1 {{hep-th/#1}}
\def\lr { \lref}
\def\npb#1(#2)#3 { Nucl. Phys. {\bf B#1} (#2) #3 }
\def\rep#1(#2)#3 { Phys. Rept.{\bf #1} (#2) #3 }
\def\plb#1(#2)#3{Phys. Lett. {\bf #1B} (#2) #3}
\def\prl#1(#2)#3{Phys. Rev. Lett. {\bf #1} (#2) #3}
\def\physrev#1(#2)#3{Phys. Rev. {\bf D#1} (#2) #3}
\def\ap#1(#2)#3{Ann. Phys. {\bf #1} (#2) #3}
\def\rmp#1(#2)#3{Rev. Mod. Phys. {\bf #1} (#2) #3}
\def\cmp#1(#2)#3{Comm. Math. Phys. {\bf #1} (#2) #3}
\def\mpl#1(#2)#3{Mod. Phys. Lett. {\bf #1} (#2) #3}
\def\ijmp#1(#2)#3{Int. J. Mod. Phys. {\bf A#1} (#2) #3}
\def\mpla#1(#2)#3{Mod. Phys. Lett. {\bf A#1} (#2) #3}
\def\jhep#1(#2)#3{JHEP {\bf  #1} (#2) #3}

\parindent 25pt
\overfullrule=0pt
\tolerance=10000

\def\half{{\textstyle {1 \over 2}}}
\def\thalf{{\textstyle{{3\over 2}}}}


%
\lr\polchinski{J. Polchinski, 
       {\it Dirichlet branes and Ramond-Ramond charges}, 
 \prl75(1995)4724, \xxx9510017.}
\lr\bounstrefa{C.G. Callan, C. Lovelace, C.R. Nappi and S.A. Yost,
  {\it Adding holes and crosscaps  to the superstring},
  \npb293(1987)83.  }
\lr\bounstrefb{J. Polchinski and Y. Cai, {\it Consistency of open
    superstring theories}, \npb296(1988)91.}
\lr\gepner{D. Gepner, {\it Space-time supersymmetry in compactified
    string theory and superconformal models}, \npb296(1988)757; {\it
    Exactly solvable string compactifications on manifolds of $SU(N)$
    holonomy}, \plb199(1987)380.}
\lr\reck{A. Recknagel and  V. Schomerus, {\it D-branes in Gepner models},
   \npb531(1998)185, \xxx9712186.} 
\lr\cardy{J.L. Cardy, {\it Boundary conditions, fusion rules and the
   Verlinde formula}, \npb324(1989)581.}
\lr\cardylew{J.L. Cardy and D.C. Lewellen, {\it Bulk and boundary
    operators in conformal field theory}, \npb259(1991)274.}
\lr\mgys{M.Gutperle and Y. Satoh, {\it D-branes in Gepner models and
supersymmetry}, \xxx9808080.}
\lr\ishibashi{N. Ishibashi, {\it The boundary and crosscap states in
    conformal field theories}, \mpla4(1989)251.}
\lr\oogurioz{ H.~Ooguri, Y.~Oz and Z.~Yin, {\it D-Branes on Calabi-Yau
 spaces and their mirrors}, \npb477(1996)407,  \xxx9606112. }
\lr\wittenbc{E.~Witten, {\it  Chern-Simons gauge theory as a string theory}, 
 \xxx9207094.}
\lr\affleck{I. Affleck, {\it Boundary condition changing operators in
    conformal field theory and condensed matter physics}, \xxx9611064.}
\lr\wittentop{E. Witten, {\it Mirror manifolds and topological field
    theory}, Nucl. Phys. Proc. Suppl. {\bf 58} (1997) 35, \xxx9112056.}
\lr\ademodels{J. Fuchs, A. Klemm, C.  Scheich and  M.G. Schmidt, 
{\it Spectra and symmetries of Gepner models compared to
    Calabi-Yau compactifications}, \ap204(1990)1.} 
\lr\blackpbrane{G.T. Horowitz and A. Strominger, {\it Black strings
    and p-branes}, \npb360(1991)197.}
\lr\devech{P. Di Vecchia, M. Frau, I. Pesando, S. Sciuto, 
     A. Lerda and R. Russo, {\it Classical p-branes from boundary
    state}, \npb507(1997)259, \xxx9707068.}
\lr\wittena{E. Witten, {\it Phases of N=2 theories in 
 two-dimensions}, \npb403(1993)159, \xxx9301042.}
\lr\sagnotti{G. Pradisi,  A. Sagnotti and Y.S. Stanev, {\it
    Completeness conditions for boundary operators in 2-D conformal
    field theory}, \plb381(1996)97.}
\lr\polchinski{J. Polchinski,  {\it Tasi lectures on D-branes}, \xxx9611050.}  
\lr\billo{M. Bill{\'o}, P. Di Vecchia, M. Frau, A. Lerda, I. Pesando, 
      R. Russo and S. Sciuto, {\it Microscopic string analysis of the D0 -
    D8-brane system and dual R - R states}, \npb526(1998)199,  \xxx9802088.}
 \lr\bbs{K.~Becker, M.~Becker and  A.~Strominger, {\it Fivebranes,
 membranes and non-perturbative string theory}, 
  \npb456(195)130, \xxx9507158. }
\lr\bb{K. Becker and  M. Becker, {\it Instanton action for type II
hypermultiplets}, \xxx9901126.}
\lr\Lelw{D.C. Lewellen, {\it Sewing constraints for conformal field
theories on surfaces with boundaries}, \npb372(1992)654.}
\lr\sagnottib{G. Pradisi, A. Sagnotti and  Y.S. Stanev, {\it
Completeness conditions for boundary operators in 2-D conformal field
theory }, \plb381(1996)97, \xxx9603097.}
\lr\fuchs{J. Fuchs and C. Schweigert, {\it Branes: from free fields to
    general backgrounds}, \npb530(1998)99, \xxx9712257; ~{\it
    Classifying algebras for boundary conditions and traces on spaces
    of conformal block}, \xxx9801191.}
\lr\greengut{M.B. Green and M. Gutperle, {\it Light-cone supersymmetry and
               D-branes}, \npb476(1996)484, \xxx9604091.}
\lr\BFIS{M. Bertolini, P. Fr{\'e}, R. Iengo and C.A. Scrucca,
         {\it Black holes as D3-branes on Calabi-Yau threefolds}, 
          \plb431(1998)22, \xxx9803096.}
\lr\ntwosugra{A. Ceresole, R. D'Auria and S. Ferrara, 
             {\it The symplectic structure of N=2 supergravity 
                  and its central extension}, 
          Nucl. Phys. Proc. Suppl. {\bf 46} (1996) 67, \xxx9509160; 
            ~L. Andrianopoli, M. Bertolini, A. Ceresole, R. D'Auria, 
             S. Ferrara, P. Fr{\'e} and T. Magri, 
             {\it N=2 supergravity and N=2 super Yang-Mills theory on 
                  general scalar manifolds}, 
       J. Geom. Phys. {\bf 23} (1997) 111, \xxx9605032. }
\lr\Sabra{W.A. Sabra, 
         {\it Black holes in N=2 supergravity
           theories and harmonic functions}, \npb510(1998)247, \xxx9704147.}
\lr\FKS{S. Ferrara, R. Kallosh and A. Strominger, 
           {\it N=2 extremal black holes}, \physrev52(1995)5412, 
              \xxx9508072.}
\lr\Behrndt{K. Behrndt, {\it Quantum corrections for D=4 black holes and 
             D=5 strings}, \plb396(1997)77, \xxx9610232.}
\lr\HINS{ F. Hussain, R. Iengo, C. N{\'u}{\~n}ez and C.A. Scrucca, 
       {\it Interaction of moving D-branes on orbifolds}, 
       \plb409(1997)101, \xxx9706186.}
\lr\Fre{P. Fr{\'e}, {\it Supersymmetry and first order equations for 
         extremal states: monopoles, hyperinstantons, black-holes and 
         p-branes}, \xxx9701054.} 
\lr\lustrew{D. L\"ust, 
 {\it String vacua with N=2 supersymmetry in four dimensions}, 
  \xxx9803072.}
\lr\larusigor{I.R. Klebanov and L. Thorlacius, {\it The Size of
    p-branes}, \plb371(1996)51, \xxx9510200.}
\lr\typeI{C. Angelantonj, M. Bianchi, G. Pradisi, A. Sagnotti
     and Y.S. Stanev,
    {\it  Comments on Gepner models and type I vacua in string
    theory}, \plb 387(1996)743, \xxx9607229; 
    ~R. Blumenhagen and A. Wisskirchen, 
    {\it Spectra of 4D, N=1 type I string vacua on non-toroidal 
     CY threefolds}, \plb438(1998)52, \xxx9806131.}
\lr\kachru{S. Kachru, A. Lawrence and E. Silverstein,  {\it On the
    matrix description of Calabi-Yau compactifications},
  \prl80(1998)2996, \xxx9712223.}
\lr\Geplect{D. Gepner, {\it Lectures on $N=2$ string theory}, in
     Superstrings '89, ed. M.B. Green et al, World Scientific, p.238-302.}
\lr\paraf{V.A. Fateev and A.B. Zamolodchikov, {\it 
    Nonlocal (parafermionic) currents in two-dimensional conformal 
    field  theory and self-dual critical points in $Z_N$ symmetric
    statistical systems}, Sov. Phys. JETP {\bf 62}(1985) 215.}
\lr\parafb{G. Mussardo,   G. Sotkov and M. Stanishkov, {\it 
    Fusion  rules, four point functions and discrete  symmetries of $N=2$
    superconformal models}, \plb218(1989)191.}
\lr\dkps{M.R. Douglas, D. Kabat, P. Pouliot
                  and S.H. Shenker, {\it D-branes and short distances
                    in string theory}, \npb485(1997)85, \xxx9608024.}
\lr\attrac{S. Ferrara and  R. Kallosh, {\it Supersymmetry and
    Attractors}, \physrev54(1996)1514, \xxx9602136.}
\lr\recktwo{A. Recknagel and V. Schomerus, {\it Boundary deformation
    theory and moduli spaces of D-branes}, \xxx9811237.}
\lr\frau{ M. Frau, I. Pesando, S. Sciuto, A. Lerda and R. Russo, 
        {\it Scattering of closed strings from many D-branes}, 
   \plb400(1997)52, \xxx9702037.}
\lr\bachas{C. Bachas, {\it D-brane dynamics}, \plb374(1996)37, \xxx9511043.}
\lr\BG{ O. Bergman and M.R. Gaberdiel, 
  {\it A non-supersymmetric open string theory and S-duality}, 
    \npb499(1997)183, \xxx9701137.}
\lr\dzeroactn{M. Bill{\'o}, S. Cacciatori, F. Denef, P. Fr{\'e},
   A. Van Proeyen and D. Zanon, 
   {\it The 0-brane action in a general D=4 supergravity background}, 
    \xxx9902100.} 
\lr\BCV{M. Bill{\'o}, D. Cangemi and P. Di Vecchia, 
        {\it Boundary states for moving D-branes}, 
         \plb400(1997)63, \xxx9701190.}
%
%
\noblackbox
\baselineskip 14pt plus 2pt minus 2pt
\baselineskip 20pt plus 2pt minus 2pt
\Title{\vbox{\baselineskip12pt
\hbox{hep-th/9902120}
\hbox{PUPT-1838}
}}
{\vbox{
\centerline{D0-branes in Gepner models and $N=2$ black holes} }}

\centerline{Michael Gutperle\foot{email:
gutperle@feynman.princeton.edu} and 
Yuji Satoh\foot{email: ysatoh@viper.princeton.edu}} 
\bigskip
\baselineskip 17pt plus 2pt minus 2pt
\centerline{Department of Physics, Princeton University} 
\centerline{Princeton, NJ 08544, USA} \bigskip

%
\vskip 1in
\centerline{{\bf Abstract }}
\bigskip
\baselineskip 17pt plus 2pt minus 2pt
In this paper D-brane  boundary states constructed in
Gepner models are used  to analyze some aspects of 
the dynamics of D0-branes in Calabi-Yau  compactifications 
of type II theories to four dimensions. It is shown that 
the boundary states correspond to  BPS objects carrying 
dyonic charges. By analyzing the couplings to closed string 
fields a correspondence between the D0-branes and extremal 
charged black holes in $N=2$ supergravity is found.

\noblackbox
\baselineskip 14pt plus 2pt minus 2pt

\Date{February 1999}
\vfill\eject
\newsec{Introduction}
Dirichlet branes \polchinski\  provide a remarkably simple way to
introduce  extended objects carrying Ramond-Ramond (RR) charges in 
string perturbation theory. One way to describe D-branes is given by
the boundary state formalism \bounstrefa\bounstrefb. Here the open string
boundary conditions are imposed on the closed string fields. The 
boundary state formalism is particularly useful in determining the
coupling of the D-brane to the various closed string modes.

The compactification of type II string theories on Calabi-Yau 
threefolds (CY) yields  four-dimensional theories with $N=2$ spacetime  
supersymmetry. This is an interesting field of study  
since these theories are not as strongly constrained as theories with more 
supersymmetry. In addition  they exhibit
many of the phenomena which are central to the recent developments in
non-perturbative string theories such as  mirror symmetry, F-theory and
heterotic duals (see \lustrew\  for a review).

The wrapping of D$p$-branes on $p$-dimensional supersymmetric cycles 
leads to BPS saturated non-perturbative objects. From the  point of 
view of the four-dimensional non-compacct space 
these objects are  D0-branes, i.e. particle like states. 
For type IIB on a CY, D3-branes can wrap
3-cycles  and  for type IIA on a CY, D0,D2,D4 and D6 branes can wrap
0,2,4 and 6 cycles respectively. Note that the NS five-brane which is 
present in both IIA and IIB cannot give any particle like states 
in four dimensions. The NS five-brane 
will however lead to non-perturbative corrections to the hypermultiplet
geometry via euclidean wrapping on the whole CY \bbs\bb.

As we shall see the D0-branes are BPS objects which are charged 
with respect to the  RR-vector fields. One aim of this paper is 
to show that they carry the same charges as extremal charged black  
holes in the $N=2$ supergravity theory defined by the low energy limit  
of the type II string theory compactified on the CY. In the context of
compactifications on $T^6/Z_3$ orbifolds this correspondence has been
discussed in \BFIS. 

Gepner models \gepner\ are exactly solvable models for strings
compactified on CY manifolds, where the internal $c=9$ SCFT is
constructed as an orbifold of a tensor product of $N=2$ minimal
models. Note that the geometrical description in sigma-model language
and the Landau-Ginzburg phase in which the Gepner model resides  can be
connected via a linear sigma-model \wittena. It would  be interesting
to study wrapped D-branes in the context of the linear sigma-model.

The construction of boundary conformal field theories from bulk
conformal field theories has been discussed by various people
\ishibashi\sagnotti. Boundary CFT has important applications 
for impurity problems  in condensed matter physics \affleck. 
The construction of consistent (rational) boundary conformal field
theories has been discussed by Cardy \cardy.  
Additional sewing constraints for such theories have been discussed 
in \Lelw\sagnottib. Cardy's boundary states have been used
in \reck\ to construct boundary states in the Gepner model and some
aspects of these boundary states were discussed further in \mgys\foot{
Gepner models in type I theory context were discussed in \typeI.}.

In this paper we will use boundary states in Gepner models to discuss
the dynamics of D0-branes in four dimensions. This analysis will be
done by applying the well known techniques developed for  D-brane
boundary states in ten-dimensional  Minkowski space to the Gepner model
compactifications. 

\newsec{Boundary states in Gepner models}
The construction of boundary states for free bosons and fermions
satisfying Neumann or Dirichlet boundary conditions 
uses coherent states \bounstrefa. This construction was generalized
for rational CFT's by Ishibashi \ishibashi. 
When the conformal field theory forms an extended algebra, boundary
conditions have to be specified on the left- and right-moving 
 currents of the extended algebra $W,\bar{W}$ \ishibashi. 
To construct  boundary states for rational conformal field
theories  one first defines Ishibashi states  
$ \mid  i  \,  \rangle \rangle $ 
for every primary field defining an irreducible highest weight
representation ${\cal H}_i$ of the algebra  
which satisfy \ishibashi\ 
\eqn\symalg{\big( W_n-(-1)^{h_W}\bar{W}_{-n}\big)\mid i \, \rangle \rangle = 0
\period }
 In \ishibashi\  it was shown that an Ishibashi
state can be constructed using an anti-unitary operator $U$ which 
acts on the modes of the right-moving current $\bar{W}$ in the following way,
$U \bar{W}_n U^{-1}= (-1)^{h_W} \bar{W}_n$.
Such an operator $U$ is closely related to the chiral CPT operator. 
Explicit form of the Ishibashi state is given by
\eqn\ishibashia{\mid i \, \rangle\rangle= \sum_N \mid i,N \, \rangle \otimes
U\widetilde{\mid i,N \, \rangle   \comma }}
where $N$ denotes the sum over the basis of ${\cal{H}}_i$.
In the second step a boundary state can be 
constructed from a complete set of Ishibashi states, 
\eqn\bounstatea{\mid a  \, \rangle= \sum_i B^a_i \mid i
  \, \rangle\rangle \period }
The form of possible boundary states $\mid a \, \rangle $ 
are constrained by the fact that the cylinder amplitude involving 
two  boundary states is mapped into a one-loop open string partition 
function by a modular transformation. Cardy \cardy\ gives a solution 
in terms of the modular S-matrix $B^a_i =S^a_i/\sqrt{S^a_0}$.

In \oogurioz\wittenbc, it was shown that two different boundary
conditions for $U(1)$ current $J$ and the  superconformal generators
$G^\pm$ are consistent with  $N=1$ superconformal invariance.
The two cases are called  A and B boundary conditions, referring to the two
possible topological twists of the $N=2$ theory \wittentop. The A
boundary conditions are defined by
\eqn\Atypebc{(J_n-\bar{J}_{-n})\mid B \, \rangle=0,\quad (G^\mp_r+i\eta
  \bar{G}^\pm_{-r})\mid B \, \rangle=0 \comma }
whereas the B boundary conditions are  defined  by
\eqn\Btypebc{(J_n+\bar{J}_{-n})\mid B \, \rangle=0,\quad  (G^\pm_r+i\eta
  \bar{G}^\pm_{-r})\mid B \, \rangle=0 \period }
The choice of $\eta=\pm  1$ corresponds to a choice of spin structure. 
Due to the property of the operator $U$ the states  \ishibashia\
satisfy B-type boundary conditions. In order to implement A-type boundary
conditions $U$ has to be replaced by $U\Omega$   \oogurioz\reck,
where $\Omega$ is the mirror  automorphism of the $N=2$ algebra.

For four-dimensional compactifications preserving $N=2$ spacetime
supersymmetry the internal
degrees of freedom form a $c=9$ $N=2$ SCFT.  In the light-cone gauge
the  external physical degrees
of freedom are given by a $c=3$ $N=2$ SCFT consisting of a free
complex boson and fermion. The boundary
conditions in the light-cone gauge have been discussed in \mgys.  
The results we need here are:  Firstly D0-brane
boundary conditions correspond to A-type boundary conditions for the
external CFT and secondly the 
GSO projection relates the internal and external boundary conditions
where the consistent internal boundary conditions for D0-branes are 
A-type  for type IIB  and B-type for  type IIA. 

Gepner models are exactly  soluble string backgrounds in which the
internal $c=9$ SCFT is an orbifold of a tensor product of $N=2$
minimal  models. Appendix A gives a brief overview on the most
important facts needed in this paper.
In \reck\  boundary states in Gepner models corresponding to total 
A and B boundary conditions were constructed by applying Cardy's construction
 \cardy\ for each factor  of the tensor product of $n$ minimal $N=2$
theories.   

The Ishibashi states are labelled by the quantum numbers for  the
external part and the internal $N=2$ primaries, which we denote by
$\mu = (s_0; m_1,\cdots,m_n; s_1,\cdots,s_n)$ and $\lambda =
(l_1,\cdots,l_n)$. For A-type boundary conditions for each 
minimal model, the Ishibashi states have the form $\mid
\lambda,\mu;\; \lambda, \mu \, \rangle\rangle$.  Whereas for B-type
boundary conditions the Ishibashi states have the form $\mid
\lambda,\mu;\; \lambda, -\mu \, \rangle\rangle$. 

In the following we will focus on the A-type boundary conditions for
the internal $c=9$ SCFT, which corresponds to D0-branes in type IIB. 

A boundary state is then labeled by a vector $\alpha=
(\lambda^\prime,\mu^\prime)$ where 
$\lambda^\prime=(l^\prime_1,\cdots,l^\prime_n)$ and $\mu^\prime=
(s^\prime_0;m_1^\prime,\cdots,s^\prime_n)$, and given  by
\eqn\boundstate{\mid\alpha \, \rangle = {1\over \kappa_\alpha}
  \sum^\beta_{\lambda,\mu}   
B^{\alpha}_{\lambda,\mu}\mid \lambda,\mu \, \rangle\rangle \period}
Although every boundary state in \boundstate\ satisfies
A boundary conditions,  different $\alpha$ gives 
different boundary conditions, e.g., for the supercharges.
The normalization constant ${1/ \kappa_\alpha}$ 
can be determined by Cardy's condition \reck.
The factor $B^{\alpha}_{\lambda,\mu}$ is the product of
$B^a_i$   
 using the modular $S$-matrix for the $N=2$ minimal models  \reck;
\eqn\bsol{B^{\alpha}_{\lambda,\mu}= e^{i\pi s_0^2/2}e^{-i\pi
    {s_0s_0^\prime\over 2}}
  \prod_{j=1}^N{\sin\big(\pi{(l_j+1)(l^\prime+1)\over k_j+2}\big)\over
   \sin^{1/2}\big(\pi{(l_j+1)\over k_j+2}\big)}
  e^{i\pi{m_jm^\prime_j\over k_j+2}}e^{-i\pi {s_js^\prime_j\over 2}
\period }}

The boundary states \boundstate\ are constructed using two types of
data, firstly the `gluing automorphism' \reck (i.e. $U$ and  $U\Omega$),
and secondly the set of $B^\alpha_{\lambda,\mu}$. It would be
interesting to explore the possible generalizations of these boundary
states. One possibility involves 
more general automorphisms of the chiral fusion algebra (a very simple
example are permutations of minimal models of the same level
$k$). Another possibility is to  construct  more general
$B^\alpha_{\lambda,\mu}$ than Cardy's. In particular, in the correspondence of
D-brane boundary states in Gepner models and CY-compactifications, the
boundary conditions in \Atypebc\ and \Btypebc\ are only necessary for
the currents of 
the $c=9$ conformal field theory, not for the currents associated with
each minimal model factor in the  tensor
product. Relaxing this condition makes the theory non-rational and more
general boundary states are possible \foot{We are grateful to
  A. Recknagel  and V. Schomerus  for correspondence on this subject.}.
A general solution to  this problem seems rather difficult  and  we
will not  pursue this  question further in this paper (see however \fuchs).
\newsec{Covariant external part of the boundary state}
The construction reviewed in the previous section is purely 
algebraic in terms of $N=2$ SCFT. However, the external part 
consists of usual free bosons and fermions. Thus one can split the boundary 
states into the internal and external parts and express the 
external part by coherent-type boundary states. 
In such an expression, the degrees
of freedom from the boson zero-modes become explicit.

In the following we will construct the external  part of the boundary
state in a covariant formalism following \devech.  In the light-cone
gauge all longitudinal degrees of freedom are expressed in terms of
transverse ones. It is possible to formulate D-brane boundary states
in the light-cone gauge \greengut  (see appendix B). There are however
some aspects of the boundary states which are best described in the
covariant Ramond-Neveu-Schwarz formalism. In the covariant formalism,
in addition to the bosonic
and fermionic world sheet fields and internal CFT fields, there are
anti-commuting $b,c$ ghosts 
and commuting $\beta,\gamma$ ghosts. The boundary conditions for the
ghosts are determined by demanding BRST invariance of the boundary
state. The main feature of the ghost dependent part is 
that it cancels the contribution of two bosonic and fermionic 
oscillators in the annulus partition function, leading to the same
result as the light-cone boundary states. There are however
subtleties arising in the RR sector due to the choice of a picture for
the superghosts, which will be addressed later.

A D0-brane imposes Neumann  boundary conditions on $X^0$  and
Dirichlet boundary conditions  on
$X^i,i=1,2,3$. The boundary conditions for the bosonic oscillators are
given by
\eqn\onebranebc{\big(a_n^\mu- M^\mu_\nu \bar{a}^\nu_{-n}\big)\mid
  B \, \rangle_X=0 \comma }
where $M^\mu_\nu=diag(-1,+1,+1,+1)$. 
The D0-brane will be localized in space at $y^i,i=1,2,3$. The boundary
state is then given by 
\eqn\xboundst{\mid B \, \rangle_X = {T_0 \over 2} \int 
  {d^3p \over (2 \pi)^3} 
   \ e^{-ip_i y^i} \exp\big( -
  \sum_{n>0}{1\over n}M_{\mu\nu}a_{-n}^\mu \bar{a}_{-n}^\nu\big)\mid
  p^i \, \rangle \comma}
with $ T_0 $ a normalization constant. This is determined so that 
the boundary state reproduces the cylinder amplitudes from 
the Coleman-Weinberg formula \polchinski. For D$p$-branes in $D$-dimensions, 
it is given by \frau\devech, 
\eqn\Tp{
   T_p = { \sqrt{\pi} \over 2^{(D-10)/4}} (4 \pi^2 \alpha')^{(D-2p-4)/4}
   \period
} 
The D0-brane tension is then  
given by $ T_0/ \kappa $ where $ \kappa^2 = 8 \pi G_N $ with 
$ G_N $ Newton's constant.

The $b,c$ part of the boundary state will not be important in the
following and  will not be displayed here.
In the NSNS sector the world-sheet fermions are half integral modes
and the boundary condition on the fermions and $\beta,\gamma$ ghosts
are given, e.g., by
\eqn\onebrbctwo{\big( \psi_r^\mu -i\eta
M^\mu_\nu\bar{\psi}^\nu_{-r}\big)\mid B,\eta \, \rangle_\psi=0 \comma}
with 
\eqn\zerobrbc{\mid B,\eta \, \rangle_\psi= \exp\big( i\eta
  \sum_{r>0}(M_{\mu\nu}\psi_{-r}^\mu\bar{\psi}_{-r}^\nu+\gamma_{-r}\bar{\beta}_{-r}-\beta_{-r}\bar{\gamma}_{-r})\big)\mid 0 \, \rangle_{(-1,-1)} \period}
Here the subscript $(-1,-1)$ indicates that in the NSNS sector the
ghost number of the vacuum is chosen to  be $(-1,-1)$.
The boundary state \boundstate\ are  GSO projected, 
which means that the total U(1) charge of the left- as well as 
the right-moving sector is odd. The boundary state \zerobrbc\ has 
neither even nor odd fermion number. There are two possible combinations of  
definite fermion number, 
\eqn\boundstsz{\eqalign{\mid B, s_0=0 \, \rangle_{NSNS}
   &={1\over 2}\big( \mid   B,+1 \, \rangle_\psi+\mid   B,-1 \, 
  \rangle_\psi\big)\cr
   &= \cos\big(\sum_r (
   M_{\mu\nu}\psi^\mu_{-r}\bar{\psi}^\mu_{-r} 
   +\gamma_{-r}\bar{\beta}_{-r}-\beta_{-r}\bar{\gamma}_{-r})\big)
   \mid 0 \, \rangle_{(-1,-1)}} \comma} 
and 
\eqn\boundstsz{\eqalign{\mid B, s_0=2 \, \rangle_{NSNS}&={1\over 2}\big
( \mid   B,+1 \, \rangle_\psi-\mid   B,-1 \, \rangle_\psi\big)\cr
  &= i \sin\big(\sum_r (
  M_{\mu\nu}\psi^\mu_{-r}\bar{\psi}^\mu_{-r} 
  +\gamma_{-r}\bar{\beta}_{-r}-\beta_{-r}\bar{\gamma}_{-r})\big)
   \mid 0 \, \rangle_{(-1,-1)}} \period}
Here $\mid B, s_0=0 \, \rangle_{NSNS}$ has even fermion
number and has to be tensored with a state with odd  internal $U(1)$
charge. On the other hand  $\mid B, s_0=2 \, \rangle_{NSNS}$
has odd fermion number and has to be tensored with a state with even
internal $U(1)$ charge.

In the RR sector the implementation of the GSO projection is more subtle. 
The ground states in the RR sector are tensor
products of spin fields. Due to the superghost number anomaly on the
disk the boundary state is not defined in a left-right symmetric
picture but in the $(-3/2,-1/2)$ picture. There are two components 
labeled by $s_0=\pm 1$ which have external $U(1)$ charges. 
In analogy with the ten-dimensional case described in \billo\ 
we can construct the RR part of the boundary state as a product
\eqn\Rten{\eqalign{ & \mid B, s_0=+1 \, \rangle_{RR}
   = \cos\big(\gamma_0\bar{\beta}_0 \!+\! \sum_{n>0}M_{\mu\nu}\psi_{-n}^\mu
  \bar{\psi}^\nu_{-n} \!+\! \gamma_{-n}\bar{\beta}_{-n} \!-\!
  \beta_{-n}\bar{\gamma}_{-n}\big) \ \sigma^0_{a\dot{b}}  \mid
  a \, \rangle_{-\thalf}\mid \dot{b} \, \rangle_{-\half}\cr
 & \qquad \qquad \qquad + i \sin\big( \gamma_0\bar{\beta}_0 \!+\! 
  \sum_{n>0}M_{\mu\nu}\psi_{-n}^\mu
  \bar{\psi}^\nu_{-n} \! + \!  \gamma_{-n}\bar{\beta}_{-n} \!-\!
  \beta_{-n}\bar{\gamma}_{-n}\big) \ \bar{\sigma}^0_{\dot{a}{b}}  \mid
  \dot{a} \, \rangle_{-\thalf}\mid {b} \, \rangle_{-\half}} \comma }
and 
\eqn\Releven{\eqalign{& \mid B, s_0=-1 \, \rangle_{RR}
  = \cos\big(\gamma_0\bar{\beta}_0 \!+ \! \sum_{n>0}M_{\mu\nu}\psi_{-n}^\mu
  \bar{\psi}^\nu_{-n} \! + \! \gamma_{-n}\bar{\beta}_{-n} \! - \!
  \beta_{-n}\bar{\gamma}_{-n}) \ \sigma^0_{\dot{a}{b}}  \mid
  \dot{a} \, \rangle_{-\thalf}\mid {b} \, \rangle_{-\half}\cr
& \qquad \qquad \qquad +  i\sin\big(\gamma_0\bar{\beta}_0 \! + \! 
   \sum_{n>0}M_{\mu\nu}\psi_{-n}^\mu
  \bar{\psi}^\nu_{-n} \! + \!  \gamma_{-n}\bar{\beta}_{-n} \!- \!
  \beta_{-n}\bar{\gamma}_{-n}\big) \ \bar{\sigma}^0_{{a}\dot{b}}  \mid
  {a} \, \rangle_{-\thalf}\mid \dot{b} \, \rangle_{-\half}} \period}
Here $\mid a \, \rangle_s, \mid \dot{a} \, \rangle_s $ denote four-dimensional
spinor fields in the $s$-picture (see  appendix C for details on  the
covariant description of the  RR fields in this context). 

The GSO projection implies that the boundary state $\mid B,
s_0=-1 \, \rangle_{RR}$ is tensored with states in the internal sector
$\mid  q,q \, \rangle$ with $q= 2m+1/2, m\in Z$ whereas  $\mid B,
s_0=+1 \, \rangle_{RR}$ is tensored with states in the internal sector
$\mid  q,q \, \rangle$ with $q= 2m-1/2, m\in Z$. 

We can now calculate the external contribution to the cylinder
amplitude. Denoting the contributions from the boson zero-modes and 
the boson and fermion oscillators by $\chi_0 $ and $ \chi_{s_0}$ respectively, 
it is given by
\eqn\extcyl{\eqalign{
  & \langle \, B, s_0, y_1 \mid q^{\half(L_0+\bar{L_0}-c/12)}\mid B,
 s_0, y_2 \, \rangle = \chi_0 \chi_{s_0} \comma \cr 
 &  
  \chi_0 = V_0 \Bigl( {T_0 \over 2} \Bigr)^2 { t^{-3/2} \over (2 \pi)^3 }
   \ e^{ -(y_1-y_2)^2/(4 \pi t) } \comma \ \  
  \chi_{s_0} = \eta^{-2} \left\{ \matrix{  
  {1\over 2 } \bigl[ {\theta_3\over  \eta}+ (-1)^{s_0/2} 
  {\theta_4 \over  \eta} \bigr] 
  &  { \rm for } \; s_0=0,2 \cr
 {1\over 2} {\theta_2 \over  \eta} &  {\rm for} \;s_0=\pm 1 } \right.
   \comma }}
with $q = e^{2\pi i \tau } = e^{- 2\pi t }$ and $ V_0 $ the world-line volume.
The combinations of theta functions are nothing but the $SO(2)$
characters  $\chi_o,\chi_v,\chi_{s/c}$ for $s_0=0,2,\pm1$ respectively. 
\newsec{Open string partition function}
A distinct advantage of knowing the exact conformal field theory and
the boundary states is that 
the  closed string cylinder amplitude has a dual interpretation
as the open string partition function. The two are related by
world-sheet duality transformation  which mixes all the massless and the
massive modes. 
 
For A boundary conditions the cylinder amplitude for two boundary
states implementing the same boundary conditions $\alpha$ is given by
\eqn\gepmdcyl{ Z_{\alpha\alpha}(q)= \langle \, \alpha \mid 
   q^{ {1\over2} (L_0+\bar{L}_0-c/12) } \mid \alpha \, \rangle
 = { 1\over \kappa^2_\alpha}
   \sum_{\lambda,\mu}^\beta
  B^\alpha_{\lambda,\mu} B^\alpha_{\lambda,-\mu}\chi^{\lambda}_\mu(q) 
\period }
Here we have suppressed the contribution from the boson zero-modes 
$\chi_0$.
A modular transformation into the open string channel gives 
\eqn\modgeptr{Z_{\alpha\alpha}(\tilde{q})=  
 { 1\over \kappa^2_\alpha}  \sum_{\lambda,\mu} ^\beta
\sum^{\rm ev}_{\bar{\lambda},\bar
     {\mu}}B^\alpha_{\lambda,\mu}
   B^\alpha_{\lambda,-\mu}S^{\lambda,\mu}_{\bar
{\lambda},\bar{\mu}}\chi^{\bar{\lambda}}_{\bar{\mu}} (\tilde{q}) \comma}
where $ \tilde{q}= e^{- 2 \pi/t} $ and 
$\sum^{\rm ev}$ stands for the constraints $ l_i + m_i + s_i = 2 Z $.
This expression can be evaluated using the explicit form of
$B^\alpha_{\lambda,\mu}$ \bsol\ and the modular matrix $S^{\lambda,\mu}_{\bar
{\lambda},\bar{\mu}}$ for the Gepner models. This calculation was done
in \reck. Here we only need the result with the same boundary conditions on
both ends of the cylinder, 
\eqn\reckres{Z_{\alpha\alpha}(\tilde{q}) =  
\sum_{\bar{\lambda},\bar{\mu}}^{\rm ev} 
\sum_{v_0=0}^{K-1}\sum_{v_1,\cdots,v_n=0,1}(-1)^{\bar{s}_0}
\delta^{(4)}_{\bar{s}_0,2+v_0+2\sum v_i}
\prod_{j=1}^{n} N_{l^\prime_j}^{\bar{l}_j}
\delta^{(2k_j+4)}_{\bar{m}_j,v_0}\delta^{(4)}_{\bar{s}_j,
v_0+2v_j}\chi^{\bar{\lambda}}_{\bar{\mu}}(\tilde{q}) \period }
Here the condition that the characters
$\chi^{\bar{\lambda}}_{\bar{\mu}}$ in \reckres\ appear only in
integer multiplicities determines the normalization $1/\kappa^2_\alpha$
up to an overall integer factor.   
$ \delta^{(k)}_{m,n} $ are non-zero for $ m= n$ (mod $k$). 
$N^{l_2}_{l_1}$ is the matrix appearing in the fusion rules
among the primaries with spin $l_{1,2}/2 $ in the  $SU(2)_k$ WZW model; 
$ \phi_{l_1/2} \times \phi_{l_1/2} 
\sim \sum_{l_2} N^{l_2}_{l_1} \phi_{l_2/2} $. Namely, $ N^{l_2}_{l_1} = 1 $ 
for $ 0 \leq l_2 \leq \min (2l_1, 2k-2l_1) $ and otherwise vanishing.  
Note that the open string partition function  $Z_{\alpha\alpha}$ for two
identical D-branes only depends on 
$\lambda=(l^\prime_1,\cdots,l^\prime_n)$ in
$\alpha=(\lambda^\prime,\mu^\prime)$.
It is easy to see that the open string spectrum of \reckres\ 
is GSO projected, i.e. the all the states appearing 
in $Z_{\alpha\alpha}(\tilde{q})$ have
odd integer $U(1)$ charges. However the set of labels
$ (\lambda,\mu) $ for fields entering  in
the open string partition function \reckres\ is very different from the
fields entering  the closed string boundary state \gepmdcyl,
since all  $m_i=v_0$ in  \reckres\ take the same value .

The primary fields appearing in \reckres\ define a set of boundary
operators $\psi^{(\alpha\alpha)}_{\lambda,\mu}$. Here the label
$ (\alpha\alpha)$ stands for operators which do not change the
boundary condition $\alpha$. In general there are also `boundary
condition changing' operators $\psi^{(\alpha\beta)}$, which
inserted at a point on the boundary will change the boundary
condition from $\alpha$ to $\beta$. The spectrum of such
operators can be determined considering
$Z_{\alpha\beta}(q)=\langle \, \alpha\mid \Delta \mid \beta \, \rangle$
instead of \gepmdcyl. For boundary conditions $\alpha,\beta$
which are not mutually supersymmetric there will be an open
string tachyon in the spectrum.  
The `massless' open string modes for D0-branes in Gepner models
correspond to primaries  with $\tilde{q}^0$ in the partition function.  

As  an example,  we consider the quintic hypersurface which corresponds to the
$(k=3)^5$ Gepner model. There is a `universal' $\psi_{\lambda_g,\mu_g}$,
with labels  $s_0=2, (l_i,m_i,s_i)=(0,0,0)^5$. This  corresponds   
to the excitations $\psi^\mu_{-1/2}\mid 0 \, \rangle, \mu =0,\cdots,3$. In
spacetime these modes correspond to the position of the D0-brane and a
gauge field $A_0$ living on the world-line. 
 
If $l^\prime_j=0,3$  the   $N^{\bar{l}_j}_{l^\prime_j}$ in \reckres\
is only non-zero for  $\bar{l}_j=0$. It is then easy to see using the
formulas given  in Appendix A that  the fields given above  are the
only massless fields if there exits $ \ l'_j = 0,3 $. 
These fields and their fermionic partners
are a $d=4$ $N=1$ vector multiplet reduced to $0+1$
dimensions. Introducing $U(n)$ Chan-Paton factors the resulting quantum
mechanics describing $n$ D0-branes will be given by the dimensional
reduction  of $N=1$ $U(n)$ Super Yang-Mills to $(0+1)$ dimensions.

If all $l^\prime_j=1,2$  
there are additional massless fields denoted by $s_0=0,
(l_i,m_i,s_i)=(2,4,2)^5$ which by field redefinition is mapped into $s_0=0,
(l_i,m_i,s_i)=(1,-1,0)^5$  and $s_0=0,
(l_i,m_i,s_i)=(2,6,2)^5$ which is mapped into $s_0=0,
(l_i,m_i,s_i)=(1,1,0)^5$. For the quintic one can check that these two
are the only massless states (together with their fermionic partners)
and they correspond to scalars (fermions) in the external sector.
Hence depending on the Gepner model and the boundary conditions
$\alpha$ there are additional massless modes which are given by the
dimensional reduction of a chiral $N=1$ multiplet to $(0+1)$
dimensions. For $n$ D0-branes this multiplet would  transform in the
adjoint of  $U(n)$. In principle the couplings of these fields are
determined by calculating correlation functions on the disk with
$\psi^{\alpha\alpha} $ operators inserted on the boundary.
The simplest such quantity would be given by the
three-point function of two fermionic open string states and one
bosonic one, which is connected to the superpotential of the chiral fields.

When there is more than one type of D0-brane present there can also be
additional massless (and tachyonic) modes coming from open strings
stretched between the two branes. Such open string states are given
by boundary changing operators $\psi^{\alpha\beta}$.

It is an interesting and open question whether the  quantum mechanics
of D0-branes in compactified theories can be used to give a Matrix
theory definition of M-theory compactified on CY \kachru.
In this context, D0-branes in type IIA are relevant and they 
are described by the B-type boundary states.
In this case, the expression of the cylinder amplitude is
slightly different from \reckres\ \reck. Then one finds that 
if all $l'_j = 1,2 $ the characters including massless states are 
the same in the closed and open channels, which is similar 
to the ten-dimensional case.
\newsec{Coupling to closed string fields}
Boundary states for D$p$-branes can be efficiently used \devech\ to
calculate the coupling (tadpoles) of massless closed string modes to the
D-brane. Such couplings were then compared to the large
distance behavior of fields around a black $p$-brane of
supergravity \blackpbrane. 

In the case of D0-branes in Gepner model compactifications the boundary
state will be used to read  off the coupling to the gravitational field
and the vector multiplets.

The massless states in the NSNS sector corresponding to the universal
fields are of the form
\eqn\univerf{\mid \xi_{\mu\nu} \, \rangle = \xi_{\mu\nu}\psi_{-\half}^\mu
\bar{\psi}_{-\half}^\nu\mid k  \, \rangle \comma }
where the four-dimensional polarization tensor contains the graviton,
NSNS anti-symmetric tensor and dilaton. 
The dilaton polarization tensor is given by
$\xi^\phi_{\mu\nu}=1/\sqrt{2}(\eta_{\mu\nu}-n_\mu  k_\nu-n_\nu k_\mu)$
where the vector $n_\mu$ satisfies $n\cdot  k=1,n^2=0$. 
The coupling of the dilaton to the D0-brane vanishes for D0-brane boundary 
conditions defined in
\onebranebc\ since $\xi^\phi_{\mu\nu} M^{\mu\nu}=0$. This result is
expected since there is no direct coupling of the D0-brane to the
hypermultiplet moduli. The coupling of the anti-symmetric tensor
also vanishes.
Using the traceless-ness of the graviton the coupling of the  graviton
to the D0-brane boundary state is given by
\eqn\gravcoupl{\langle \, \xi_{\mu\nu} \mid
 B \, \rangle= M^{\mu\nu}\xi_{\mu\nu} \calN B^{\alpha}_{\lambda_g,\mu_g}
 =\xi_{00} \calN B^{\alpha}_{\lambda_g,\mu_g} \period}
Here $\calN = T_0/(2\kappa_\alpha)$,  
we have defined $\lambda_g$ by $(l_i = 0)$ and
$\mu_g$ by $(s_0=2,m_i=s_i=0)$ and in the last line the traceless-ness 
of the graviton wave function has  been used.
 
The scalar  fields in  the vector multiplets are found in the NSNS
sector. Each scalar field corresponds to a primary
$\Phi^{\lambda,\bar{\lambda}}_{\mu,\bar{\mu}}$  with $h=\bar{h}=1/2$.
For type IIB, which we are interested in,   
we have $q=\pm 1,\bar{q}=\pm1$ whereas for type IIA  we have
$q=\pm 1,\bar{q}=\mp1$. 
We denote by $\phi^A$ a particular scalar  
associated with a primary $\Phi^{\lambda,{\lambda}}_{\mu,{\mu}}$ with
$q=1$, and by $\bar{ \phi}^A$ a conjugate field with primary
$\Phi^{\lambda,{\lambda}}_{-\mu,{-\mu}}$ and $q=-1$. The one-point
function for the two scalars is then  given by
\eqn\sacltadp{\langle \,  \lambda,\mu, \;
\lambda,\mu\mid B \, \rangle = \calN 
 B^\alpha_{\lambda,\mu}, \quad \langle \,  \lambda,-\mu, \;
\lambda,-\mu\mid B \, \rangle = \calN  B^\alpha_{\lambda,-\mu}
= \calN  B^{\alpha\;*}_{\lambda,\mu} \comma}
where the last equality follows from the definition of
$B^\alpha_{\lambda,\mu}$ \bsol.

For the scalar field $\phi^A$ or the graviton 
with quantum numbers $(\lambda,\mu)$, 
there is a  vector field in the Ramond sector associated with $s_0=+1$   
and it has the quantum numbers $(\lambda,\mu-\beta_0)$. 
Then the charge conjugated component
with $s_0=-1$ has quantum numbers  $(\lambda,-\mu+\beta_0)$. 
Because the
RR boundary states  \Rten\ and \Releven\ are in the $(-3/2,-1/2)$
picture the relevant tadpoles use the vector vertex operators
$W^{\pm}$ which are discussed in Appendix C. In particular, 
it turns out that only the components $W^\pm_0(A) $  
couple to the boundary states. 
The relevant part of the boundary state \boundstate\ is  given by
\eqn\vectad{\mid B \, \rangle = \cdots + B^\alpha_{\lambda ,\mu-\beta_0}
  \mid \lambda,\mu-\beta_0 \, \rangle+ B^\alpha_{\lambda,-\mu+\beta_0}\mid
  \lambda,-\mu+\beta_0 \, \rangle \comma}
where the  external part in the above is given by \Rten\ or
\Releven\ depending on the value of $s_0=\pm 1$.
As discussed in Appendix C, the couplings of the boundary states 
to $W^\pm_0 (A) $ are electric and magnetic, respectively. 
They are determined by 
using the explict forms of the external boundary
state \Rten\ and \Releven\ and the vertices $W^\pm$,  
\eqn\Rthirteen{\langle \, W_0^+(A)\mid B \, \rangle= \calN 
\big(B^\alpha_{\lambda,\mu-\beta_0}
+B^\alpha_{\lambda,-\mu+\beta_0}\big)A_0^e, \ \ 
   \langle \,  W^-_0(A)\mid B \, \rangle= \calN 
 \big(B^\alpha_{\lambda,\mu-\beta_0}-B^\alpha_{\lambda,-\mu+\beta_0}\big)
 A_0^m\period}
Hence the boundary states are dyonic and 
the electric and magnetic charges are given by 
\eqn\emcharge{q_e^{i}=\calN \left| B^\alpha_{\lambda,\mu-\beta_0}\right| 
 \cos \theta \comma  \quad  q_m^{i}= \calN 
 \left| B^\alpha_{\lambda,\mu-\beta_0} \right| \sin \theta \comma }
where the angle $\theta$ depends on the quantum numbers of 
the vector field and the boundary condition $\alpha$.
The graviphoton field has quantum numbers $\lambda= \lambda_g$
and $\mu = \mu_g$.
Thus using the formula for $ B^\alpha_{\lambda,\mu} $ \bsol, 
the angle $ \theta_0 $ is given by 
\eqn\anglgravph{\theta_0 
 = 2 \pi \mu^\prime \bullet \beta_0 
 =-\pi{s_0^\prime\over 2}+\pi \sum_{i=1}^r\Big(
{m_i^\prime \over k_i+2}-{s_i^\prime\over 2}\Big) = \pi Q_\alpha 
  \comma}
where $Q_\alpha $ defines a 
$U(1)$ charge associated with the boundary
condition $\alpha$ (see Appendix A). Note that the condition for  mutually
supersymmetric branes is given by $ Q_{\alpha}=
Q_{\tilde{\alpha}}+2n,n\in Z$.
For the vector field associated with a scalar $\phi^A$ 
with $(\lambda,\mu)$, the angle $\theta_A$ is given by 
\eqn\angelaph{\theta_A 
  = -\pi{s_0^\prime  s_0\over 2}+\pi\sum_{i=1}^r \Big({m_im_i^\prime\over
    k_i+2}-{s_is_i^\prime\over 2}\Big)-\pi Q_\alpha \period}
Alternatively for a given boundary condition $\alpha$ one might 
introduce a phase in the definition of the
vertex operators $W^\pm_{0}$ and make the coupling to the boundary state
purely electric or magnetic. For different boundary conditions
$\tilde{\alpha}$  the  coupling will  in general  be dyonic\foot{
These results are obtained also by a simple argument in the light-cone
gauge. In this case, it is easy to calculate 
$ \bra{ A_1 \pm i A_2 } \semiket{ B } $. After an analytic continuation
(see Appendix B), these give the couplings of $ A_0 \pm i A_2 $. 
However, since the gauge fields are supposed to live on the world-line, 
the field strength 
coming from $ A_2 $ should be interpreted as a magnetic dual of 
a field strength, namely,  $ F_{\mu \nu } (A_2) = * F_{\mu \nu} (A'_\mu) $
where $ A'_\mu $ has only $A'_0$ component. Then one finds the same dyonic 
coupling as in \emcharge.    
}.

Note that a connection between $(l_i,m_i,s_i)$ and the geometrical picture 
of the boundary states has been discussed in \mgys.
In this sense, the ratio of the electric and magnetic charges may 
be determined by geometrical data similarly to the orbifold limit 
case \BFIS.  

In ten dimensions D-brane charges automatically satisfy 
Dirac-Zwanziger quantization conditions \polchinski. 
In the case of CY compactifications  there are $n_V + 1$
different vector fields and charges, and for two different boundary
conditions $\alpha$ and $\tilde{\alpha}$ we denote the charges by $q$ and
$\tilde{q}$ respectively. The quantization condition should then read 
\eqn\diracqu{q_e^0\tilde{q}_m^0
  -q_m^0\tilde{q}_e^0+\sum_{A=1}^{n_V} 
 \Big( q_e^A\tilde{q}_m^A -q_m^A\tilde{q}_e^A\Big)
 = 2\pi n,  \quad n\in Z \period}

Hence we would like to interpret the boundary state as a D0-brane
which carries electric and magnetic charges $q_e,q_m$ and has a
following coupling to the world-line to the lowest order in the bulk fields;
\eqn\worldlineact{S= \int d\tau (1+ c_A^* \phi^A+ c_A
  \phi^{A*})\sqrt{g_{\mu\nu}\dot{x}^\mu\dot{x}^\nu }+ \big(q^0_e A^0_\mu 
  +q^0_m A^{0\;D}_\mu+q^A_e A^A_\mu 
  +q^A_m A^{A\;D}_\mu\big) \ \dot x^\mu \period}
Here the  values for $c_i$ are determined by \sacltadp\ and the
charges  $q^i$ are determined by \emcharge.  
\newsec{Supersymmetry and vanishing of cylinder amplitude}
The BPS nature of D0-branes defined by the boundary states
\boundstate\ ensures that the static force  between  two of them
vanishes. This force can be calculated to  lowest order in string
perturbation theory by the cylinder diagram. 

In the limit of large separations only the exchange of massless
closed string states is important. The vanishing of the cylinder
amplitude implies the BPS cancelation between the massless NSNS and
RR fields. We will see that this  occurs 
in each multiplet. In the  universal sector the cancelation between the
graviton and the graviphoton exchange determines the BPS relation
between the mass and the central charge. For the vector multiplets
the cancelation relates the scalar and the vector coupling.

The cancelation between the NSNS and RR fields at every mass level 
follows from two facts. Firstly the contribution of 
a state to the cylinder amplitude is proportional to 
\eqn\bbisca{\eqalign{
  C_{\lambda,s_0}^{\lambda'} &=
   B^\alpha_{\lambda, \mu} B^\alpha_{\lambda, -\mu} \cr
    &= (-1)^{s_0} \prod_{j=1}^n \sin^2 
  \Bigl[ \pi {(l_j+1)(l_j'+1)\over k_j+2} \Bigr]
   \Bigm/ \sin \Bigl[\pi {l_j+1\over k_j+2} \Bigr] \period} }
Secondly, the supersymmetry ensures that 
for every state in the NSNS sector there is an RR state in the 
same mass level.
The supersymmetry connecting the NSNS and RR states is  
essentially the spectral flow and shifts $\mu$ in \bbisca\ 
by an odd number of $\beta_0$. Since $\beta_0$ is independent 
of $\lambda, \lambda'$, the contributions from the NSNS and RR states 
have just an opposite sign,  
\eqn\cnsrra{
 C_{\lambda,s_0=\pm 1}^{\lambda'} = - C^{\lambda'}_{\lambda,s_0=0,2} 
  \period}
This gives the cancelation in each multiplet at each mass level.

Alternatively, we can apply an argument in the light-cone gauge 
given by Gepner in \Geplect.
The spacetime supersymmetry current  can be bosonized  defining (see
Appendix B for details on the notation)
\eqn\totbos{\phi
={1\over 2}\phi_{ext}+ {\sqrt{3}\over 2} H \period}
The exponentials by $\psi= e^{i \phi}$
and $\psi^* =e^{-i\phi}$ (omitting a cocycle factor) are free
fermions. The fact that all fields have odd $U(1)$ charge implies that
the fermion $\psi $ is in the Ramond sector (this is not to be confused
with the NSNS and  RR sector described above). 

The supersymmetry charge is  simply  the zero-mode of these free
complex fermions  $Q=\psi_0,Q^\dagger=\psi_0^*$. 
The $c=12$ SCFT  splits into a $c=1$
free complex fermion and a $c=11$ CFT, which is neutral under the $U(1)$
charge.  Acting on a state with $Q$ will
shift $\mu\to \mu+\beta_0$. 

Similarly for the right-mover we have a free complex fermion
$\bar{\psi}$ associated with the supercharges
$\bar{Q},\bar{Q}^\dagger$. 
In \mgys\ it was shown that the boundary state  
preserves  a  supersymmetry $(Q+i\bar{Q}^\dagger)\mid
B \, \rangle=0$. This  implies that   the boundary state must be 
the product of a $c=11$ part times  
\eqn\zermq{\mid B \, \rangle _{c=1} 
  \ \sim \ \mid --\, \rangle \,  + i \psi_0^*\bar{\psi}_0^* \mid -- \, 
  \rangle \comma}
where $ \mid -- \, \rangle $ is annihilated by 
$\psi_0$ and $\bar{\psi}_0$. 
It is clear from \bbisca\ that a shift by
$\beta_0$ introduces a minus sign for each pair $
\psi^*_0 \bar{\psi}^*_0$ inserted. Hence the cylinder amplitude will
vanish since    $\langle\,  -- \mid -- \, \rangle \, - \, \langle \, ++ \mid
 ++ \, \rangle \,  = 1-1 = 0 $ with $ \mid ++ \, \rangle$ a state 
annihilated by $ \psi_0^* $ and $ \bar{\psi}_0^* $.   
After modular transformation to the open string channel, the same
argument basically implies that the free fermion part of the partition
function is give by $\tr(-1)^F q^{L_0}=0$, which vanishes because of
the presence of fermionic zero modes for  
the free fermion $Q$ in the odd spin structure.   
This is reminiscent of the GS formalism where the Green-Schwarz
fermions become world-sheet fermions in the light-cone gauge, and
cancelations due  to  
supersymmetry  are  manifest in the odd spin structure. 
\newsec{Velocity dependent cylinder amplitude}
The boundary state formalism can be used to describe D-branes
moving with a constant velocity \BCV. Such boundary states are simply
constructed by applying a boost operator on the boundary states at
zero velocity. Hence the boosted boundary state is  defined by
\eqn\rotbound{
   \ket{ B, v } \equiv 
  e^{iv K^{01}} \ket{ B, 0 }  \comma} 
where $\tanh v $ is the velocity and  
the boost operator $K^{01}=K^{01}_X+K^{01}_\psi$ is given by 
\eqn\boostopa{K^{01}_X= x^0p^1-x^1p^0 -i \sum_{n>0}{1\over n}
 (a_{-n}^0  a_{n}^1 - a_{-n}^1 a_{n}^0 + \bar{a}_{-n}^0
  \bar{a}_{n}^1 - \bar{a}_{-n}^1 \bar{a}_{n}^0 ) \comma}
and the fermionic part in the NSNS and RR sector respectively
\eqn\boostopb{\eqalign{K^{01}_\psi&= -i\sum_{r>0} 
 (\psi_{-r}^0 \psi_{r}^1 - \psi_{-r}^1 \psi_{r}^0 + 
  \bar{\psi}_{-r}^0 \bar{\psi}_{r}^1 -
  \bar{\psi}_{-r}^1 \bar{\psi}_{r}^0 )\quad ({\rm NSNS}) \comma \cr
K^{01}_\psi&= -{i\over 2}
 ([\psi_0^0,\psi_0^1]+[\bar{\psi}_0^1,\bar{\psi}_0^0])
 -i\sum_{n>0} (\psi_{-n}^0 \psi_{n}^1 - \psi_{-n}^1 \psi_{n}^0 
 + \bar{\psi}_{-n}^0 \bar{\psi}_{n}^1 -
  \bar{\psi}_{-n}^1 \bar{\psi}_{n}^0 ) \;\; ({\rm RR}) \period} }
It is easy to see that as far as the cylinder amplitude is concerned
the boost is just acting as a twist on the bosonic and fermionic
oscillators. After careful  consideration of the zero-mode part the
changes in the cylinder amplitude can be separated into the bosonic
and fermionic part and we get
\eqn\chartwist{\eqalign{
\chi_0 \quad & \to \Bigl( {T_0 \over 2} \Bigr)^2 { t^{-1} \over (2 \pi)^2 }
      \ e^{- b^2 /(4 \pi t) } \comma \cr
\chi_{s_0=0,2} & \to {1\over i \theta_1 (iv/\pi \vert it) } 
  \Bigl[ \theta_3(iv/\pi|it) + (-1)^{s_0/2} \theta_4(iv/\pi|it) \Bigr] 
   \comma \cr
\chi_{s_0=\pm 1} &\to {1\over i \theta_1 (iv/\pi \vert it)} 
  \theta_2(iv/\pi|it) \comma 
}}
where $ b^2 = \sum_{i = 2,3} (y_1^i -y_2^i)^2 $.
\subsec{potential between two moving D0-branes}
For non-zero $v$, the cylinder amplitude does not vanish and
one obtains non-vanishing potential between two moving D0-branes. 
The calculation is straightforward and, by the standard procedure, 
we get
\eqn\potential{ 
  V(r) = -  {1 \over \kappa^2_\alpha} \int_0^\infty {dt \over t^{1/2}}
    \ \chi_0 \bigm\vert_{b \to r} \sum_{\lambda, \mu}^\beta 
    C^{\lambda'}_{\lambda, s_0} \chi^\lambda_\mu (q)
   \comma
}
where $r$ is the separation of the two D0-branes and 
$\chi_{s_0}$ in $ \chi^\lambda_\mu $ are given by \chartwist.

The contributions from the massless closed string states 
are extracted using the asymptotic forms of the characters 
as $ q = e^{-2\pi t} \to 0$. At low velocity, the leading 
contributions are given by
\eqn\ptmasles{
  V (r) = - {\calN^2  \over 2 \pi r} 
   \Bigl[ C^{\lambda'}_{\lambda_g, 2} (\cosh 2v - \cosh v) 
         + \sum_c C^{\lambda'}_{\lambda, 0} (1 - \cosh v) \Bigr]
 \period
}
Here  
$ \sum_c $ indicates the summation over the chiral primaries and
includes the contributions from $n_V$ vector multiplets whereas
the first term is the contribution from the gravity multiplet.
We see that the cancelation at $v=0$ occurs in each multiplet.

We also see that the potential starts from $ v^2 $ and hence the 
moduli space has a non-trivial metric. This is as expected
because the boundary states preserve only $N=1$ spacetime supersymmetry. 
Moreover, since theta functions $ \theta_i(iv/\pi\vert it)$ can be expanded 
in $ v $ by Eisenstein functions, it is possible to get the 
exact coefficients for $v^n$ terms.
\subsec{absorptive part} 
At non-zero $v$, the cylinder amplitude 
${\cal A} = (\pi/2) \int_0^\infty dt \  Z(q) $ gets imaginary part
because $ \theta_1^{-1} $ has simple poles in the open string 
channel. The imaginary part is interpreted as indicating pair creation of 
open strings as discussed in the ten-dimensional case \bachas. 
Although the amplitudes in Gepner models are more complicated, 
the argument here is almost the same as in the ten-dimensional case 
since we know the modular transformation between the closed and 
open string channels. 
 
Following the procedure in \bachas, the imaginary part is given by
\eqn\imA{ 
   {\rm Im} {\cal A} = { \calN^2 \over 4 \pi } \sum_{k=1}^\infty
    {1 \over k } \ e^{- k b^2/(4v) } \chi({k \pi \over v })
  \period
}
Here $ \chi (s) = (i \theta_1/2 \eta^3) Z_{\alpha \alpha} (\tilde{q}) $
with $\tilde{q} = e^{-2\pi s} $ and 
$Z_{\alpha \alpha} (\tilde{q})$ 
is given by \reckres\ and $ \chi_{s_0}$ in \chartwist\
with $\theta_i (is v/\pi \vert is)$ instead of 
$\theta_i (i v/\pi \vert i t)$.

At low velocity $ v \ll 1$, the leading contribution comes from 
the massless open sting states and it is vanishingly small for $b^2 > v $
as $ e^{-b^2/v} $. 
On the other hand, at high velocity $ v \gg 1$, the dominant
contribution comes from the graviton in the closed string channel
and the internal minimal model part does not matter. By a modular
transformation of $\chi(s)$, one finds 
\eqn\imAuv{
   {\rm Im} {\cal A} \sim {\calN^2 \over 4 } v^{-1}
   e^{- b^2/(4v) + v } C^{\lambda'}_{\lambda_g, 2}
  \period 
}      
This shows the behavior of fundamental strings at high energy
and black absorptive disks with area $ \ln (s/M_0^2) $ where
$s$ and $M_0$ are the center of mass energy squared and the 
$ D0$-brane mass, respectively. 

\newsec{Comparison to black holes in $N=2$ supergravity}
D-branes in the ten-dimensional theory have corresponding 
black $p$-brane solutions in supergravity \blackpbrane.
They are regarded as different representations of the same
BPS objects. Since D-branes have boundary state description,
it should be possible to relate them to the black $p$-branes.
In fact, a precise correspondence has been found in this case
\devech\greengut. Furthermore, a four-dimensional Reissner-Nordstrom 
black hole is obtained by wrapping a D3-brane 
on a Calabi-Yau threefold and 
the connection to the corresponding boundary state has been 
shown in the orbifold limit $T^6/Z^3$ \BFIS.     
In this section, we show that our boundary states correspond
to some extremal black holes in $D=4$ $N=2$ supergravity. 

Our boundary states couple to the bosonic sector of 1 gravity
multiplet and $n_V = h^{2,1}$ vector multiplets, where $h^{2,1}$
is a Hodge number of the corresponding CY. Note that there are no 
couplings to the hypermultiplets (including the dilaton) and 
the anti-symmetric tensor. The boundary states also satisfy the 
BPS condition. 

In $D=4$ $N=2$ supergravity, there exist the BPS black hole solutions
which have the same couplings. To see this, let us first recall 
the low energy action for the bosonic part of the 
gravity and vector multiplets\foot{
For a review of $D=4$ $N=2$ supergravity, see, e.g., 
\ntwosugra.
}
\eqn\action{
  S =  \int \sqrt{-g} \ d^4x \lbb \half R - g_{A\bar{B}} \del^\mu z^A 
     \del_\mu z^{\bar{B}} 
        -  \calF^I_{\mu \nu} (\ast \calG_I^{\mu \nu})
        \rbb \comma 
}
with
\eqn\calg{
     \calG_{I \, \mu \nu} = {\rm Re} \calN_{IJ} \calF^{J}_{\mu \nu} - 
                    {\rm Im} \calN_{IJ} \ast{\cal F}^{J}_{\mu \nu}
    \period
}
Here the complex scalars $ z^A $ $(A= 1, ..., n_V)$ parametrize 
a special K{\" a}hler manifold with the metric 
$ g_{A \bar{B}} = \del_A \del_{\bar{B}} K(z,\zbar) $ where $ K $ is 
the K{\"a}hler potential. The symplectic period matrix $ \calN_{IJ}$
$ (I,J = 0, ..., n_V) $ and the K{\"a}hler potential are expressed by the
holomorphic potential $ F(X)$ :
\eqn\NK{\eqalign{
   K &= - \log \lbb i (\bar{X}^I F_I - X^I \bar{F}_I) \rbb \comma \cr
   \calN_{IJ} &= \bar{F}_{IJ} + 2 i 
  {({\rm Im}F_{IK}) X^K ({\rm Im}F_{JL}) X^L \over X^M ({\rm Im}F_{MN}) X^N} 
   \comma
}}  
with $ F_I = { \del F \over \del X^I } $ and 
$ F_{IJ} = {\del^2 F \over \del X^I \del X^J}$. $X^I$ are homogeneous 
coordinates and satisfy the constraint 
$  L^I {\rm Im} F_{IJ} \bar{L}^J = -1/2$ with $ L^I = e^{K/2} X^I$. 
In Calabi-Yau compactifications, $F(X)$ is given by
\eqn\FX{  F(X) = d_{ABC} {X^A X^B X^C \over X^0}
  \comma
}
where $d_{ABC}$ are the intersection numbers of the CY.
The physical field strengths of the graviphoton
and the other vector fields are given by
\eqn\tg{
   T^-_{\mu \nu} = 2 i {\rm Im}\calN_{IJ} L^I \calF^{- J}_{\mu \nu} 
   \comma \qquad
   G^{- A}_{\mu \nu} 
    =  - g^{A \bar{B}} \bar{f}^I_{\bar{B}} 
        {\rm Im}\calN_{IJ} \calF^{-J}_{\mu \nu}
  \comma
 } 
where $ f^I_A = (\del_A + \half \del_A K) L^I $
and $-$ stands for the antiself-dual part such as 
$\half (T_{\mu \nu} - i \ast T_{\mu \nu})$.
The electric and magnetic charges are defined by 
\eqn\em{
   q_I = { 1 \over 4 \pi} \int_{S^2_\infty} \calG_{I \, \mu \nu}
         d x^\mu \wedge d x^\nu \comma \qquad 
   p^I = { 1 \over 4 \pi} \int_{S^2_\infty} \calF^I_{\mu \nu}
         d x^\mu \wedge d x^\nu 
   \period
}
The complex charges of $ \calF^{-I}_{\mu \nu}$, which are defined 
similarly to \em, are related to $ q_I $ and $ p^I $ by 
\eqn\qpt{
    t^I =  {1 \over 2} \Bigl[ p^I + i ({\rm Im}\calN^{-1})^{IJ} 
        \lb {\rm Re}\calN_{JK} p^K - q_J \rb \bigm\vert _{\infty} \Bigr]
   \period
}
If we define the electric and magnetic charges by the asymptotic 
behavior
\eqn\emtwo{
   \calF^I_{0i} \sim {e^I \over r^3} x^i \comma \qquad
    \ast\calF^I_{0i} \sim {g^I \over r^3} x^i \comma 
}
with $ r^2 = x^i x_i $ $(i =1,2,3)$, they are related to $t^I$
by  
\eqn\teg{
   t^I = {1 \over 2} ( g^I - i e^I ) \period 
} 

Here we consider the spherically symmetric solutions whose 
metric can be written in the form
\eqn\metric{ 
   ds^2 = - e^{2U}(r) dt^2 + \ e^{-2U}(r) \lb dr^2 + r^2 d\Omega_2^2 \rb
  \period  
} 
The BPS solutions without non-trivial fermionic fields are obtained 
from the condition of vanishing supersymmetry transformations for 
the gravitino and the gauginos, 
$ \delta \psi_{\alpha \mu } = \delta \lambda^{A \alpha } = 0 $.
For a particular choice of the supersymmetry parameter, 
these conditions give the first order differential equations \Fre,
\eqn\eom{\eqalign{
   {d U \over dr} &= -2i {\rm Im}\calN_{IJ} L^I t^J {e^U \over r^2}
    \comma \cr
   {d z^A \over dr} &= -2i g^{A \bar{B}} \bar{f}^I_{\bar{B}}
     {\rm Im}\calN_{IJ} t^J {e^U \over r^2}
   \period
} }
The most general BPS black hole solutions to these 
equations are given by \Sabra
\eqn\bhsol{\eqalign{
   & z^A = X^A/X^0 \comma \qquad 
   e^{-2U} = \ e^{-K} = i (\bar{X}^I F_I - X^I \bar{F}_I ) \comma  \cr
& \calF^{I}_{ij} =  \half \epsilon_{ijk} \del_k H^I \comma \qquad
      \calG_{I \, ij} = \half \epsilon_{ijk} \del_k N_I 
      \qquad (i,j,k = 1,2,3) \comma \cr
   & i (X^I - \bar{X}^I) =   H^I \comma \qquad 
        i (F_I - \bar{F}_I) = N_I 
  \comma
}}
where $H^I(r)$ and $N_I(r)$ are the harmonic functions
\eqn\HN{
   H^I = h^I + {p^I \over r} \comma \qquad N_I = n_I + {q_I \over r}
 \period
}
In addition, we have the constraints
\eqn\cnstrnt{
  e^{-K} \vert_{r=\infty}  =  1 \comma \qquad 
 h^I q_I - n_I p^I  = 0 
 \period
}
The electric components of the gauge fields are obtained from 
\calg.
We see that these black holes have the same couplings to the 
massless fields as the boundary states.
Moreover, in a generic case, $ \calF^I_{\mu \nu}$  
are dyonic and hence so are the physical fields 
$ G^A_{\mu \nu} $. This is in agreement with 
the dyonic gauge couplings discussed in section 5.
Without introducing an additional phase in the definition of 
$ T^-_{\mu \nu} $, it follows from \tg\ and \eom\ that the 
charge of $ T^-_{\mu \nu} $ is real. Hence $ T_{\mu \nu} $ 
is always magnetic. This is again consistent with the discussion 
in section 5; by proper definition of the vertex operator,
one can always make the graviphoton magnetic 
as long as the mutually supersymmetric
boundary states, which satisfy $ Q_{\alpha} = Q_{\tilde{\alpha}}$ (mod 2), 
are considered.
 
One can find more precise correspondence using the 
potential from the massless particle exchange 
between two moving black holes.
The coupling constant (charge) for the graviton is given by
the black hole mass $M$, which is read off from the asymptotic
behavior  $ -e^{2U} + 1 \sim 2 G_N M /r $. 
Similarly, the charges for 
the complex scalars are obtained from $ z^A \sim w^A / r  + $constant.
For the gauge fields, the electric and magnetic charges are obtained
similarly to \emtwo,\teg\ and we denote by $ u^0 $ and $ u^A $ the complex
charges for $ T^-_{\mu \nu} $ and $ G^{-A}_{\mu \nu}$, respectively.
Then the potential is given by \HINS
\eqn\Vone{
 V = -{1 \over 2 \pi r} 
   \Bigl( \hat{M}^2 \cosh 2v - \abs{ \hat{u}^0 } ^2 \cosh v 
     + \sum_{A=1}^{n_V} \abs{ \hat{w}^A } ^2 
   - \sum_{A=1}^{n_V} \abs{ \hat{u}^A } ^2 \cosh v \Bigr)
 \comma
}
where $r$ is the distance between the two black holes and 
$ \tanh v $ is the relative velocity.
A hatted quantity is equal to the corresponding un-hatted one up to 
a constant coming from the proper normalization of the physical field.

Because of supersymmetry, there are constraints among charges. 
First, from the first equations of \tg\ and \eom,
one finds that $ M \sim u^0  $. This is nothing but 
the BPS mass formula for $N=2$ black holes \ntwosugra\FKS.
Furthermore, the second equations of \tg\ and \eom\ give
the relation $ w^A   \sim  i u^A  $.
Consequently, the potential is brought into the form
\eqn\Vtwo{
   V = -{1 \over 2 \pi r} \Bigl[ \hat{M}^2 (\cosh 2v - \cosh v) 
     + \sum_{i=1}^{n_V} \abs{ \hat{u}^A } ^2 (1- \cosh v) \Bigr]
  \period
}
We find that this is the same form as \ptmasles\ from Gepner models. 
The boundary state then corresponds to the black hole with 
\eqn\charges{
 \hat{M} =  \hat{u}^0  = \abs{ q_e^{0} + i q_m^{0} } 
  \comma \qquad 
  \abs{ \hat{w}^A } = \abs{ \hat{u}^A } 
  = \abs{ q_e^{A} + i q_m^{A} } 
  \period
} 

Furthermore, one can make precise comparison  
also by using the tadpoles discussed in section 5 \devech.
The couplings between the boundary states and the massless states
are regarded as the sources for the massless fields.
Thus overlaps calculated in section 5
correspond to the charges of the massless fields on the supergravity side
which are read off from the asymptotic behavior of the fields 
as $ r \to \infty$.  
However, the massless contribution to the
potential takes the form
$ V \sim r^{-1} \sum_{i: {\rm massless}} \bra{ B, v } \semiket{ i } 
\bra{ i } \semiket{ B,0 } $. 
Hence the matching of the potential 
implies the correspondence of the overlaps and the massless field charges.
In fact, this is nothing but \charges.
The advantage of using tadpoles is that one can see the tensor 
structure as in \gravcoupl\ and the dyonic couplings 
as in \Rthirteen\ \devech\BFIS. (Though we have summed the components
of the polarization tensor $\xi_{\mu \nu}$ in \gravcoupl, 
it is possible to extract a specific component, of course.) 

In addition, the boundary states can be purely magnetic (or electric).
In particular, by the proper definition of the physical 
fields, we can always make a boundary state magnetic as long as 
only one boundary state is considered.  It is easy to 
find the corresponding magnetic black hole solution on the supergravity
side. It is given by the `magnetic' solution discussed in \FKS\Behrndt, 
i.e., the solution with $ H^0 = N_A = 0 $.
In this case, the metric and the homogeneous coordinates
are given by 
\eqn\magnetic{\eqalign{
  X^0 &= {1 \over 2} \sqrt{ d_{ABC} H^A H^B H^C/ N_0}
  \comma \qquad
  X^A = - {i \over 2} H^A \comma 
  \cr
  e^{-2U} &= 2 \sqrt{ N_0 d_{ABC} H^A H^B H^C }
 \period
}}
We then find that the scalars $z^A$ are purely imaginary and so are
their charges $ w^A $. By \tg\ and \eom, this means that $ u^A $ are real
and hence $G^A_{\mu \nu}$ are magnetic. 
\newsec{Scattering involving open string states}
String scattering amplitudes on the disk with D0-brane boundary
conditions give the lowest order
contributions to scattering processes involving the D0-brane. Such
amplitudes are given by inserting  $n$  boundary
operators (open string vertices) $\psi^{\alpha\alpha}_{\lambda,\mu}$   
on the boundary of the
disk and $m$  closed string vertices
$\Phi^{\lambda,\bar{\lambda}}_{\mu,\bar{\mu}}$  in the interior 
of the disk.    

In general the boundary states which define the boundary CFT are 
determined in terms of the quantities $B^a_i$, 
where the label $i$ runs over  all the chiral 
primary fields of  the bulk CFT. The $B^a_i$ are determined via
the one-point functions of primary  fields $\Phi_{i,\Omega(i)}$ on the disk 
where  $ \langle \,  \Phi_{i,\Omega(i)} \mid \alpha  \, \rangle =
B^\alpha_i$. Here $\Omega$ denotes the `gluing automorphism' \reck\
which determines  which fields can couple consistently to the boundary.
The primary fields in the bulk live in ${\cal H}\otimes \bar{\cal H}$. 
The so called `doubling trick' turns the anti-holomorphic  field 
living on the upper half plane into a holomorphic field on the lower half 
plane,
\eqn\doupbtr{\bar{\phi}_{\Omega{(i)}}(\bar{z})\to R^\alpha_i
  \phi_{({i^+})}(\bar{z}) \comma}
where $i^+$ denotes the conjugate field which  is determined by the
chiral OPE $\phi_i(z)\phi_{i^+}(0)=z^{-2h}{\bf 1}+\cdots$.
The reflection coefficients $R^\alpha_i$ can be determined 
through the connection of
the bulk to boundary operator product
\eqn\bulkboundope{\lim_{Im(z)\to 0}\Phi_{i,\Omega{i}}(z,\bar{z})= 
  \sum_k (z-\bar{z})^{-2h_i+h_k} C^{\alpha\;k}_i \psi_k(x)
  \period
}
The OPE coefficient for the identity operator $C^{\alpha\; 0}_i$ can be
related to  the  insertion  of $\Phi_{i,\Omega(i)}$ on the disk.  
The calculation of  this quantity using the boundary states determines
$C^{\alpha\; 0}_i= B^\alpha_i/B^\alpha_0$ \cardylew. 
One can then show  that the reflection coefficient for the boundary 
conditon labeled by $\alpha$ is given by
\eqn\refcoef{R^\alpha_i= {B^\alpha_i\over B^\alpha_0} \period}

As  an example we will consider  the two-point function of two
massless closed string fields given by
\eqn\twwopoint{\eqalign{
 &  \langle \,  V_{i\Omega(j)} e^{ik_1 X}(z_1) V_{k\Omega(l)}
  e^{ik_2X}(z_2) \, \rangle_{H}  =
|z_1-z_2|^{k_1k_2}|z_1-\bar{z}_2|^{k_1Mk_2}  \cr
& \qquad \qquad \times |z_1-\bar{z}_1|^{k_1Mk_1} |z_2-\bar{z}_2|^{k_2Mk_2} 
R^\alpha_j R^\alpha_l
  \langle \,  \Phi_i(z_1) \Phi_{j^+}(\bar{z_1})  \Phi_{k}(z_2)
\Phi_{l^+}(\bar{z}_2) \, \rangle_S \comma}}
where $M$ is the matrix in \onebranebc, and 
$ \langle \cdots  \rangle_S$ denotes the correlation function for
the chiral fields on the plane.  
After fixing the conformal Killing
symmetry and integrating over the position of the vertex operators we
have a very similar result to \larusigor. The detailed properties of
the amplitude are hidden in the four-point function of the Gepner model
fields. Correlators of $N=2$ minimal models can be expressed in terms
of free bosonic and parafermionic corrrelators \paraf\parafb. 
The amplitude will display an infinite set of s-channel poles 
in the region where $z_1\to z_2$, and the bulk OPE
\eqn\bulkope{ \Phi_i(z,\bar{z})\Phi_j(0)\sim \sum_k
  z^{h_k-h_i-h_j}\bar{z}^{h_k-h_i-h_j}C_{ij}^k \Phi_k}
will determine the couplings. On the other hand  for the 
t-channel poles (when one vertex operator approaches the boundary)
the bulk to boundary OPE \bulkboundope\ determines the couplings.
The fact that these two-point functions have an infinite set
of  poles in the open string channel indicates that the D0-branes
has many internal excitations determined by open string spectrum, 
which is not surprising for a black hole.

Correlation  functions with only boundary field (open string  vertex
operators) inserted can be used to calculate terms appearing in the 
effective quantum
mechanics governing the open strings stretched between branes.  

It would be interesting to determine whether in some cases 
the massless states in the open string spectrum found in section 4
correspond to exactly marginal boundary perturbations \recktwo. 
A  condensate of such  operators would provide a continuous
deformation of the boundary state and correspond to a  modulus
associated with the wrapped brane.   
\newsec{Conclusions}
In this paper we have used the D-brane  boundary states constructed in
Gepner models to  analyze some aspects of the dynamics of D0-branes in
Calabi-Yau  compactifications of type II theories to four dimensions. 
The advantage of working with an exactly soluble conformal field
theory lies in the fact that the spectrum and in principle all the
correlation functions are exactly known. In particular this implies
that the relation between the cylinder and annulus allows to find the
spectrum of stretched open strings, which is  not possible if one only
knows the massless part of the boundary state. 

We have found a correspondence between the 
D0-brane boundary states and a class of $D=4$ $N=2$ black holes.
They are both the BPS objects for the same $N=2$ supersymmetry.
It would be interesting to find more precise correspondence
between the D0-branes and the black holes, 
e.g., the matching of the black hole entropy and the Hawking 
radiation. Gepner models and the CY compactification models
are at different points of the moduli space and the classical black holes
have macroscopic charges. Thus the issues of the moduli and quantum 
corrections might be important to further investigations.

Note that the boundary states \boundstate\ carries the minimal amount
of charge because it satisfies the Cardy's constraint with minimal
multiplicities for the open string characters. We can in principle
multiply the boundary state by an integer factor. This corresponds to
introducing $N$ Chan Paton factors. 
A disk diagram  with no open string vertices inserted simply picks 
up a trace which is equal to $N$.  Note
that by looking at a single boundary state, we  are restricted to a
regime where $N$ is finite and $g$ is small. In the ten-dimensional
case, the D$p$-branes are constructed in flat Minkowski  space, but
represent curved black $p$-branes of supergravity. The D-branes in the
Gepner model are constructed using the asymptotic vacuum defined by a
Gepner model compactification without any D-branes. On the other hand
we have seen that the BPS-character and the charges that they carry
makes it possible to identify D0-branes with extremal black holes in
$N=2$ supergravity. 

The extreme $N=2$ black holes have $AdS_2\times S_2$ near-horizon
geometries and exhibit the attractor mechanism \attrac.  
Considering a spherical world-sheet with large number of disks cut out,
i.e. a large number of boundary states on the disk, corresponds to a
planar large $N$ limit. In this limit the insertions of boundary
states should be equivalent to a new string theory on a world-sheet
with no boundaries but living in  the near-horizon geometry where the
internal Calabi-Yau manifold is determined by the fixed point moduli
of the attractor equations. 
It would be very interesting to explore
whether it is possible to connect the attractor equations with a flow
in the CFT induced by the insertion of boundaries.
\bigskip
\centerline{ {\bf Acknowledgments} }
\medskip
We are grateful to A. Recknagel and  V. Schomerus for useful
correspondence. We wish to thank  A. Lawrence, G. Lifschytz, V. Periwal,
H. Verlinde and   D. Waldram for interesting  discussions and 
conversations. We
gratefully acknowledge the hospitality of the Aspen Center for
Physics, where part of this work was done. The work of M. Gutperle was
supported by NSF grant PHY-9802484  whereas the work of Y. Satoh was 
supported by Japan Society for the Promotion of Science.
\bigskip
\noindent
{\bf Note added}
\medskip
After this work is completed, a paper appeared which discusses
related issues \dzeroactn.
\appendix{A}{Gepner models}
In this appendix we will give a brief review of some aspects of Gepner
models.  
Gepner models \gepner\ are exactly soluble supersymmetric
compactifications of
type  II and heterotic strings which use tensor products of $N=2$
minimal  models to construct the internal SCFT. The $N=2$ minimal
models  are unitary representations of the $N=2$ SCFT which are
labeled  by an integer $k=1,2,\cdots$, where the central charge is given by
\eqn\centrch{c={3k\over k+2} \period }
Primary fields  $\Phi^l_{m,s}$ are labeled by three integers $l,m,s$
with the ranges\foot{Note that states with $s=2$ are really 
descendants. Nevertheless splitting each  module into subsets with 
$s=0$ and $s=2$ is a very useful bookkeeping device. 
}
\eqn\range{l=0,1,\cdots,k,\quad  m=-(k+1),\cdots, k+2,\quad s=0,2,\pm 1 
\comma
}
together with  constraint $l+m+s\in 2Z$. The field 
identifications $(l,m,s)\sim (l,m,s+4)$
and $( l,m,s)\sim(l,m+2(k+2),s$) imply  that $m$ is defined modulo
$2(k+2)$ and $s$ is defined modulo 4. The labels $(l,m,s)$ can be
 brought into the
 `standard range' by another field identification
$(l,m,s)\sim (k-l,m+k+2,s+2)$.
The conformal dimension $h$  and $U(1)$ charge $q$
 of the primary fields (with $(l,m,s)$ in the standard rage) are given by
\eqn\confch{h = {l(l+2)-m^2\over 4(k+2)}+{s^2\over 8} \comma \quad
 q = {m\over k+2}-{s\over 2} \period }
A Gepner model is constructed by tensoring $n$ minimal models with
$k_i, i=1,\cdots, n$ such that the sum  of the central charges of the
$n$ minimal models is equal to 
\eqn\ctot{\sum_{i=1}^n{3k_i\over k_i+2}= c_{int} \period }
The total currents $T,G^{\pm},J$ of the tensor product are given 
by the sum
of the currents of each minimal model.
For $c_{int} =9 $, 
the external theory is given by one complex boson and a level one
$SO(2)$ current algebra. The primary fields can be labeled by two
vectors
\eqn\lamvec{\lambda = (l_1,\cdots,l_n),\quad
\mu=(s_0;m_1,\cdots, m_n;s_1,\cdots,s_n) \period}
Here $s_0=0,2,+1,-1$ labels the four characters corresponding to
$o,v,s,c$ conjugancy classes
 of the $SO(2)$ current algebra. 
Gepner constructed a  supersymmetric
partition function for the tensor product  by using 
`$\beta$-projections' (generalizing the GSO projection) and adding
twisted sectors to achieve  modular invariance. 
This `$\beta$-method' uses the
 $(2n+1)$-dimensional vectors:   $\beta_0$ which has $1$ everywhere  and
 $\beta_i,i=1,\cdots,n$ which has $2$ in the first and $n+1+j$ entry and 
is zero everywhere else. An inner product of two $(2n+1)$-dimensional
vectors is defined  by
\eqn\inprod{\mu \bullet \tilde{\mu}= -{1\over 4}s_0\tilde{s}_0-
\sum_{j=1}^n {s_j\tilde{s_j}\over 4}+\sum_{j=1}^n {m_j\tilde{m}_j\over
2(k_j+2)}\period}
Note that with the help of this inner product the total $U(1)$ charge
of a  primary field is given by $q_{\mu}=2\beta_0\bullet \mu$. The GSO
projection is then implemented by projecting onto states with
an odd integer charge $q_{\mu}$. In order to preserve the
$N=1$  superconformal invariance all fields in the tensor product have
to be  in the same sector (R or NS). This can be achieved by
projecting onto  states which satisfy $\beta_j\bullet \mu\in Z$ for
$j=1,\cdots n$. Gepner constructed  a modular invariant partition
function by including twisted sectors, 
\eqn\partf{Z= {1\over 2^n} ({\rm Im } \ \tau)^{-2}
    \sum_{b_0,b_j}\sum^{\beta}_{\lambda,\mu}
  (-1)^{b_0}\chi^\lambda_{\mu}(q)\chi^{\lambda}_{\mu+b_0\beta_0+\sum_jb_j   
    \beta_j}(\bar{q}) \period }
Here $b_j=0,1$;  $b_0=0,1,\cdots, K-1$;  $K= $lcm$(4,2(k_j+2))$
and $ q = e^{2\pi i \tau}$. 
$ \chi_\mu^\lambda $ are the characters corresponding to the primaries
$\Phi_\mu^\lambda$. The contribution from the external boson 
oscillators is included in $\chi_{s_0}$ in $\chi^\lambda_\mu$.
In \partf\ the diagonal affine  $SU(2)$ invariant is used which exists for
all levels  $k_j$. Other choices according to the ADE classification
of affine $SU(2)$ invariants 
 are possible and lead to different models \ademodels.
The notation $\sum^{\beta}$ indicates  the summation over 
the $\beta$-projected range $(\lambda,\mu)$ and the $(-1)^{b_0}$ imposes the
connection between spin and statistics.
Note that the supersymmetries have a very simple action on
the characters $\chi^\lambda_\mu$; acting with $Q$ corresponds to
$\mu\to \mu+\beta_0$ and acting with $Q^\dagger$ corresponds to $\mu\to
\mu-\beta_0$.
\appendix{B}{Boundary states in the light-cone gauge}
%
In \greengut, a construction of D-brane boundary states in the light-cone
gauge was  given. A peculiar feature of these states is that they
describe  $(p+1)$ D-instantons instead of D$p$-branes since the
light-cone coordinates $X^\pm$ satisfy Dirichlet boundary
conditions. One can use these boundary states simply as a calculational
tool since they are related to the D$p$-brane boundary states by an
analytic continuation (double Wick rotation). In fact, 
the calculation of cylinder amplitudes is much easier in this gauge.  

A `D$p$-brane' boundary condition is then given by 
\eqn\lcbc{
  (\del X^i \pm \bar{\del} X^i) \ket{ Bp } _X  = 0 \comma \qquad 
  ( \psi^i \pm i \eta \bar{\psi}^i ) \ket{ Bp,\eta } _\psi = 0
  \comma
}
where $i = 1,2 $, and 
plus sign for $i = 1,..., 1+p$ and minus sign for $i = 2+p, .., 2$
 $(p= \pm1,0)$.
The complex combinations of these free bosons and fermions form the simplest 
example of a $c=3$ $N=2$ SCFT.
Except for the RR zero-mode part, the boundary state is obtained 
by removing the ghost and light-cone oscillator parts of the
corresponding boundary state in the covariant case.

The construction of the RR zero-mode part is similar to 
the ten-dimensional case \BG. We start from a state $ \ket{ B1,+ } ^0_{RR}$ 
satisfying \lcbc\ for the zero-modes with $p=1$ and $\eta = + 1$.
It is useful to define 
$ \psi_\pm^i \equiv (\psi_0^i \pm i \bar{\psi}_0^i)/\sqrt{2} $ and 
$ \Gamma_3 = 2 i \psi_0^1 \psi_0^2 $, 
$ \bar{\Gamma}_3 = 2 i \bar{\psi}_0^1 \bar{\psi}_0^2$. 
Then the other zero-mode states are given by
\eqn\rrzero{ \vert \, B1, - \,   \rangle^0_{RR} 
 = \bar{\Gamma}_3 \vert \, B1, + \, \rangle^0_{RR}
  \comma \qquad  \vert \,  Bp, \pm \, \rangle ^0_{RR} 
  = \prod_{i=2+p}^2 \psi_\mp^i \vert \,  B1, \pm \, \rangle^0_{RR}
  \period
}

The GSO projection enforces that the left and right $U(1)$ charges
take odd integer values. This relates the boundary conditions of the
external and internal parts. Using free bosons, 
the left $U(1)$ current is expressed by 
$ J_{tot}= J_{ext}+J_{int}=i\partial \phi+i\sqrt{3}\partial H $
whereas the right $U(1)$ current is
\eqn\totright{{\rm IIB}: \ \
   \bar{J}_{tot}=i\bar{\partial}\bar{\phi}+i\sqrt{3}\bar{\partial}\bar{H} 
 \comma \qquad 
  {\rm IIA}: \ \  \bar{J}_{tot}=- i\bar{\partial}\bar{\phi}
   + i\sqrt{3}\bar{\partial}\bar{H} \period }
The difference between IIB and IIA is the reversal of the sign of
$\bar{\phi}$ (or equivalently the sign of $\bar{H}$). 
This can be traced back to the fact that IIB in ten dimensions
is chiral whereas IIA is not.
Because of \totright, the external chirality operators are given by 
\eqn\chiop{
  \Gamma = \Gamma_3 (-1)^F 
   \comma \qquad \bar{\Gamma} = \epsilon \bar{\Gamma}_3  (-1)^{\bar{F}}
  \comma
}
with $\epsilon = +1$ for type IIB and $-1$ for type IIA.

In A-type boundary states, 
$s_0$ takes the same sign in the left and right sectors whereas 
it takes the opposite sigh in B-type boundary states. 
Thus the GSO projected states are given by
\eqn\gsoproj{
   \ket{ Bp, s_0 = \pm 1 } = { 1 \over 2 \sqrt{2} }
   (1+s_0 \Gamma)(1 \pm s_0 \bar{\Gamma}) \ket{ Bp, + } _{RR}
  \comma
}   
where $+(-)$ for A(B)-type boundary states.
One then finds that the projected states are non-vanishing
only when (i) for type IIB, $p=0$ for A-type and $p=\pm 1$
for B-type, (ii) for type IIA, $p=\pm 1$ for A-type and $p=0$
for B-type. This is in agreement with the possible combinations
of the internal and external boundary conditions discussed in 
\mgys. This is also an analog of the ten-dimensional 
case in which  the RR boundary states exist only for odd $p$ for type IIB
and even $p$ for type IIA because of the GSO projection.     

\appendix{C}{Vertex operators in the RR sector}
The vertex operators for left- or right-moving Ramond states contain
three parts (we only display the left-moving part, the right-moving
fields will be barred):
a bosonized superghost part, an $SO(3,1)$ spin field and a Ramond
field in the internal CFT, 
\eqn\Rone{V_{s}= \exp(s\phi)\;S^A
  V_{q_{int}} \period}
Here $s$ denotes the superghost picture and $q_{int}$ denotes the
internal $U(1)$ charge. For the spin field $S^A$ we introduce a Weyl
notation where $S^a$ is a spinor  of positive helicity and $S^{\dot{a}}$
is a spinor of negative helicity. In the $s=-1/2$ picture the GSO
projected vertices are given by
\eqn\Rtwo{V^a= e^{-\half  \phi}S^a V_{q=\half,-\thalf} \comma 
  \quad  V^{\dot{a}}= e^{-\half \phi}S^{\dot{a}}V_{q=-\half,\thalf } 
    \comma }
whereas for in the $s=-3/2$ picture the GSO
projected vertices are given by
\eqn\Rthree{V_{-3/2}^{\dot{a}}= e^{-\thalf  \phi}S^{\dot{a}} 
   V_{q=\half,-\thalf} \comma \quad  V_{-3/2}^{{a}}= e^{-\thalf
    \phi}S^{{a}}V_{q=-\half,\thalf }
  \period}
This is because the GSO projection projects onto odd $U(1)$ charge where the
$U(1)$ charge is given by $q_{GSO}= s+h+q_{int}$ with $h=1$ for
un-dotted and $h=0$ for dotted spinors.

We will consider IIB strings compactified on a CY in the
following. The standard vertex operators for the RR fields are given
by tensor products of left- and right-moving RR fields in the $-1/2$
picture. These vertices contain the field strength of the appropriate
RR  potential.
This implies that for IIB the RR vector fields in the vector multiplet
are given by
\eqn\Rfour{V(F)= e^{-\half(\phi  + \bar{\phi})} e^{ikX} \Big( S^a
  \sigma^{\mu\nu}_{ab}\bar{S}^{b}\; V_{q=\half,\bar{q}=\half}
  F^+_{\mu\nu} + S^{\dot{a}}
  \bar{\sigma}^{\mu\nu}_{\dot{a}\dot{b}}\bar{S}^{\dot{b}}\; 
  V_{q=-\half,\bar{q}=-\half} F^-_{\mu\nu} \Big)       
 \period }
 Here $F^+$ and $F^-$ are the self-dual and antiself-dual part of the
 field strength, which satisfy $F^\pm= \half(F\pm i *F)$ and $*F^\pm =
 \mp i F$. For the vector multiplets in type IIB, 
 the $F^+$ part of the vertex \Rfour\  contains 
 an RR field $V_{q=1/2,\bar{q}=1/2}$ related to the (a,a) ring by
 spectral flow, whereas the $F^-$ part of the vertex \Rfour\ contains 
 an RR field $V_{q=-1/2,\bar{q}=-1/2}$ related to the (c,c) ring by
 spectral flow. These vertices correspond to the $s_0=\pm 1$ states in
 the light-cone gauge respectively. 
For the graviphoton vertex the internal part is
replaced by  $V_{q=-3/2,\bar{q}=-3/2}$ and $V_{q=3/2,\bar{q}=3/2}$
respectively which are related to the identity by spectral flow.
Because of the superghost anomaly of the disk, the RR states
which couple to D-branes are not the ones described above but are in
an asymmetric picture. This fact makes it possible that  D-branes
couple minimally to the appropriate RR-fields \sagnotti\polchinski.

The vertex operator in the $(-3/2,-1/2)$ picture is given by
\eqn\Rfive{W_0^\pm(A) = e^{-\thalf\phi -\half \bar{\phi}}
   e^{ikX} \Big( S^a 
  \sigma^\mu_{a\dot{b}}\bar{S}^{\dot{b}}
  V_{q=\half,\bar{q}=\half}\pm S^{\dot{a}} 
 \bar{  \sigma}^\mu_{\dot{a}{b}}\bar{S}^{{b}}
 V_{q=-\half,\bar{q}=-\half}\Big)A_\mu 
  \period
}
The sign will later be seen to correspond to an electric or magnetic vertex.
The picture changing operator mapping $V_{s}\to V_{s+1}$ is defined by
\eqn\Rsix{V_{s+1}(w)= \{Q_{BRST},\xi V_{s}(w)\} \comma }
where the BRST charge is  $Q_{BRST}=Q_0+Q_1+Q_2$ and the
three distinct contributions to the charge are given by
\eqn\Rnine{\eqalign{Q_0&=\oint dz \ c\big
    ( T_{ext}+T_{int}+T_{gh}\big),\quad Q_1={1\over 2} \oint
    dz \ e^{\phi}\eta\big(\psi^\mu\partial X_\mu+G_{int}\big) \comma \cr 
    Q_2&=
    {1\over 4}\oint dz \ b\eta \partial \eta e^{2\phi} \period }}
On the vertex operators of the form \Rfive\ (i.e., $s=-3/2$), 
the picture changing operation effectively reduces to
\eqn\Rsixb{V_{s+1}(w)= \lim_{z\to w} e^{\phi}G(z) V_{s}(w) \comma}
where $G= \psi^\mu \partial X_\mu +G_{int}$ and $G_{int}$ is the
internal $N=1$ supercurrent defined by $G=G^++G^-$ in terms of the two
$N=2$ superconformal currents. 
A simple observation which will be very
important in the following is that since $V_{q=\pm 1/2,\bar{q}=\pm 1/2}$ 
define Ramond/Ramond ground states  
in the internal CFT and  $V_{q=\pm 3/2,\bar{q}=\pm 3/2}$
are the internal parts of the spacetime supercharges, we
have the OPE $G(z)V_{q,\bar{q}}(w)$ which is at most $O((z-w)^{-1/2})$. 
This implies that
in the picture changing operation only the external part of \Rsix\
will be important.
Picture changing the vertex \Rfive\ gives
\eqn\Rseven{\lim_{z\to w} e^{\phi} G(z)W_0^\pm
   =   e^{-\half(\phi + \bar{\phi})} e^{ikX} \Big( S^a
  \sigma^{\mu\nu}_{ab}\bar{S}^{b} V_{q=\half,\bar{q}=\half}
  F^+_{\mu\nu} \pm S^{\dot{a}}
  \bar{\sigma}^{\mu\nu}_{\dot{a}\dot{b}}\bar{S}^{\dot{b}} 
  V_{q=-\half,\bar{q}=-\half} F_{\mu\nu}^-\Big) \comma}
where the following  OPE's were  used 
\eqn\Reight{\eqalign{\psi_\mu(z) S^a(w)&\sim {1\over
  (z-w)^{1/2}}\sigma_\mu^{a\dot{b}}S_{\dot{b}} (w),\quad 
  \partial X^\mu (z)
    e^{ikX}(w)\sim {-ik^\mu\over (z-w)} e^{ikX}(w) \comma\cr 
  &\quad
    e^{\phi}(z)e^{-\thalf\phi}(w)\sim (z-w)^{3/2} e^{-\half \phi} (w)
 \period
 }}
To derive \Rseven\  we have further to impose the Lorentz gauge condition 
$k_\mu A^\mu=0$ on the gauge potential. Note that for the picture changed 
vertex operator the choice of sign in \Rfive\  
implies that the picture changed
vertex $W^+_0$ turns into $V(F)$ with  $ dA= F$ whereas $W^-_0$ turns
into $iV(*F)$ with  $dA=*F$.  If we define the vertex $W^+_0(A^e)$
with electric variables, then the vertex $W^-_0(A^m)$ corresponds to
the magnetic dual variables.

Note that the vertex \Rfive\ in the asymmetric picture is not
BRST invariant as it stands. Due to the comment regarding the action
of $G_{int}$ above, the resolution
of this fact is exactly along the lines of the ten-dimensional case
discussed in \billo. 
The basic idea of \billo\ is that $W_0 $ fails to be BRST invariant
because of $Q_1$. A new vertex is then defined  as $W=\sum_{n=0}^\infty
W^{(n)}$ such that $[Q_1,W^{n}]+[\bar{Q}_1,W^{n+1}]=0$ for
$n=0,1,\cdots$, and hence $[Q_1+\bar{Q}_1,W]=0$. For this to work one
has to impose the Lorentz gauge condition $k_\mu A^\mu =0$. Note that
under picture changing all the $W^{(n)}$ with  $n>0$ vanish. 
\listrefs
\end